# N$_2$O weak lines observed between 3900 and 4050 cm$^{-1}$ from long path absorption spectra.

Hervé Herbin, Nathalie Picqué, Guy Guelachvili,
**Laboratoire de Photophysique Moléculaire, CNRS ; Univ Paris-Sud, Bât. 350, F-91405 Orsay Cedex, France**

Evgeni Sorokin and Irina T. Sorokina
**Institut für Photonik, TU Wien, Gusshausstr. 27/387, A-1040 Vienna, Austria**

Corresponding author:
Dr. Nathalie Picqué,
Laboratoire de Photophysique Moléculaire
Unité Propre du CNRS, Université de Paris-Sud, Bâtiment 350
91405 Orsay Cedex, France
Website: http://www.laser-fts.org
Phone nb: 33 1 69 15 66 49
Fax nb: 33 1 69 15 75 30
Email: nathalie.picque@ppm.u-psud.fr

Abstract:
Previously unobserved nitrous oxide transitions around 2.5 µm are measured by intracavity laser absorption spectroscopy (ICLAS) analyzed by time-resolved Fourier transform (TRFT) spectrometer. With an accuracy of the order of 10$^{-3}$ cm$^{-1}$, measured positions of 1637 assigned weak transitions are provided. They belong to 42 vibrational transitions, among which 33 are observed for the first time. These data are believed to be useful in particular to monitoring atmosphere purposes.

In this note, intracavity laser absorption spectroscopy (ICLAS) coupled (1) with time-resolved Fourier transform (TRFT) spectrometer is applied to the measurement of previously unobserved $N_2O$ transitions around 2.5 µm. High resolution absorption spectra of $N_2O$ in natural isotopic abundance are recorded around 4000 cm$^{-1}$ with kilometric absorption path lengths. Their analysis reveals the observed weak transitions belong to 42 vibrational transitions, among which 33 are observed for the first time. These data are believed to be useful in particular for atmospheric applications.

The laser, installed in a vacuum chamber, is made of $Cr^{2+}$: ZnSe amplifying medium inserted in X-fold cavity. It is optically pumped with an $Er^{3+}$ fiber doped laser emitting at 1.6 µm. The pumping beam is chopped by an acousto-optic modulator. The laser build-up is recorded with a TRFTS equipped with two InSb detectors cooled at 77 K. More details on the experimental setup may be found in the instrumentally oriented paper (2) and in (3) where new data measurements and analysis for 23 $C_2H_2$ molecular bands located around 4000 cm$^{-1}$ are also reported.

Three $N_2O$ time-resolved spectra (numbered 508, 509, 510) have been recorded (4) with natural sample pressures respectively equal to 3.49, 70.62, 70.58 hPa (2.62, 53.0, and 52.9 Torr). The gas was inserted in the vacuum chamber. This provides, with the elimination of the parasitic atmospheric absorption, the advantage of a laser cavity filling ratio practically equal to unity. All TRFT spectra of the laser emission were recorded with 64 time samples, 1.6 µs time resolution and 32 co-additions. Unapodized spectral resolutions and recording times were respectively 0.037, 0.037, 0.007 cm$^{-1}$ and 5, 5, and 25 minutes.

Figure 1 displays the general temporal behavior of the spectrum nb. 510 with restricted spectral resolution. The laser line is narrowing with increasing generation times reaching at most 120 µs. With 7.1 km absorption path length the spectrum covers 150 cm$^{-1}$, approximately from 3900 to 4050 cm$^{-1}$. With 33.5 km absorption path length the spectral coverage is restricted to 60 cm$^{-1}$, from 3955 to 4005 cm$^{-1}$. As in Ref.(3), it has been checked that no deviation to the linear evolution versus the generation time of the peak absorbance is observed. No attempt was made to tune the $Cr^{2+}$: ZnSe laser.

Wavenumber scale of the spectra have been calibrated against lines from residual water vapor in the laser chamber, using (5). Line positions reported here were measured in the 3 temporal samples of spectrum 510. Their sequence number is 6, 16, 26 corresponding to the equivalent absorption paths 7, 12, 17 km. Their respective wavenumber scales were first checked to be consistent within an average of 4 10$^{-4}$ cm$^{-1}$. Full width at half maximum of the $N_2O$ profiles in spectrum 510 (pressure: 70.58 hPa) is of the order of 20 10$^{-3}$ cm$^{-1}$, revealing as expected contribution of collisional broadening. The accuracy of the line position measurements is of the order of 10$^{-3}$ cm$^{-1}$. Only well-resolved and unsaturated lines were taken into account. Line sequences detected with a Loomis-Wood program (6), were left unprocessed and are not reported in this note, when at least 15 lines were not fitting the selection criteria. Altogether, 42 bands are observed. Among them, 32 belong to $^{14}N_2^{16}O$, 4 to $^{14}N^{15}N^{16}O$, 4 to $^{15}N^{14}N^{16}O$, and 2 to $^{14}N_2^{18}O$. The $^{14}N_2^{16}O$ bands consist of 1 $\Sigma$-$\Sigma$, 11 $\Sigma$-$\Pi$, 2 $\Sigma$-$\Phi$, 2 $\Pi$-$\Phi$, 4 $\Pi$-$\Pi$, 8 $\Pi$-$\Delta$ and 4 $\Delta$-$\Phi$ transitions. All the $^{14}N^{15}N^{16}O$, $^{15}N^{14}N^{16}O$, and $^{14}N_2^{18}O$ bands are $\Sigma$-$\Pi$ transitions. Practically all the observed lines are transitions between already known energy levels. Only the 13312 $^{14}N_2^{16}O$ energy level is observed for the first time. In order to appreciate the validity of our measurements, least-squares fits of individual bands were performed using the polynomial expression :

$$E(v,J) = G + BX - DX^2 + HX^3 + LX^4 \text{ with } X = J(J+1)$$

The lower state constants were held fixed in the fits to the excellent values given in (7) and (8). Only the lines measured in our spectra were processed in the fits.

The 42 observed bands are summarized with their assignments and calculated band centers in Table 1. A full resolution small part of the temporal sample n° 6 already shown in Figure 1 is given on Figure 2 with spectral assignments of the resolved lines. In the Journal supplementary material, 3 additional tables are given. Table 2 reports for $^{14}N_2^{16}O$ the effective parameters obtained from our calculations and their corresponding values taken in (7) and (8). Table 3 reports similar results for the isotopologues $^{14}N^{15}N^{16}O$, $^{15}N^{14}N^{16}O$, and $^{14}N_2^{18}O$. Table 4 is aimed to be a convenient tool. It provides in increasing order 1637 measured line positions, with their corresponding "observed – calculated" values and their isotopologue and rovibrational assignments.

The work has been supported by the French-Austrian exchange program Amadeus and the FWF project P17973.

Reference List

1. Picqué, N., Guelachvili G., Kachanov A.A., *Optics Letters* **28**, 313-315, 2003.
2. Picqué, N., Gueye, F., Guelachvili, G., Sorokin, E., and Sorokina, I. T., *Opt. Lett.* **30**, 3410-3412 (2005).
3. Girard, V., Farrenq, R., Sorokin, E., Sorokina, I. T., Guelachvili, G., and Picqué, N., *Chem. Phys. Lett.* **419**, 584-588 (2006).
4. Herbin H, *Thèse, n° d'ordre 8132, Université de Paris Sud* (2005).
5. Toth, R. A., *J. Opt. Soc. Am..* **10**, 2006-2029 (1993).
6. Brotherus, R., J. Comp. Chem. 20, 610-622 (1999).
7. Toth, R. A., *Appl. Opt.* **30**, 5289-5315 (1991).
8. Toth, R. A., *J. Mol. Spectrosc.* **197**, 158-187 (1999).

| isotop.[a] | Transition | | Band center $\nu_0$ | N lines | Observed $J_{max}$ | | | RMS |
|---|---|---|---|---|---|---|---|---|
| | upper | lower | (cm$^{-1}$) | | P | Q | R | (x 10$^4$ cm$^{-1}$) |
| 446 | 03111 | 00001 | 3931.247641(20) | 70 | 25 | - | 64 | 0.4 |
| 446 | 40001 | 02001 | 3937.54475(17) | 64 | 15 | - | 65 | 6 |
| 446 | 03311 | 00001 | 3948.28416(28) | 31 | 27 | - | 51 | 7 |
| 446 | 41101 | 03301 | 3955.91156(29) | 30 | 43 | - | 54 | 6 |
| 446 | 41102 | 03302 | 3955.91221(42) | 30 | 46 | - | 52 | 7 |
| 446 | 41101 | 03101 | 3973.75842(18) | 55 | 44 | - | 63 | 5 |
| 446 | 41102 | 03101 | 3973.75843(33) | 17 | - | 51 | - | 4 |
| 446 | 41101 | 03102 | 3973.75850(46) | 13 | - | 46 | - | 6 |
| 446 | 41102 | 03102 | 3973.75863(14) | 65 | 61 | - | 66 | 5 |
| 446 | 14011 | 03301 | 3995.46032(27) | 27 | 36 | - | 40 | 6 |
| 446 | 14212 | 03302 | 4005.71059(31) | 33 | 51 | - | 37 | 8 |
| 446 | 14211 | 03301 | 4005.71088(36) | 37 | 55 | - | 37 | 9 |
| 446 | 22011 | 11102 | 4007.83943(31) | 21 | - | 47 | - | 6 |
| 446 | 22011 | 11101 | 4007.84067(27) | 28 | 48 | - | 43 | 6 |
| 446 | 14011 | 03102 | 4013.30713(25) | 18 | - | 51 | - | 4 |
| 446 | 14011 | 03101 | 4013.30735(15) | 42 | 51 | - | 38 | 5 |
| 446 | 13111 | 02202 | 4023.03477(69) | 16 | - | 31 | - | 6 |
| 446 | 13112 | 02202 | 4023.03632(22) | 25 | 48 | - | 10 | 4 |
| 446 | 13112 | 02201 | 4023.03636(43) | 17 | - | 34 | - | 6 |
| 446 | 13111 | 02201 | 4023.03638(29) | 35 | 41 | - | 21 | 5 |
| 446 | 14212 | 03102 | 4023.55756(24) | 31 | 53 | - | 20 | 5 |
| 446 | 14211 | 03101 | 4023.55819(15) | 29 | 64 | - | 16 | 4 |
| 446 | 13112 | 02001 | 4032.64887(37) | 28 | - | 55 | - | 5 |
| 446 | 13111 | 02001 | 4032.64889(20) | 47 | 61 | - | 8 | 5 |
| 446 | 21111 | 10001 | 4034.27055(17) | 46 | 66 | - | 2 | 5 |
| 446 | 21112 | 10001 | 4034.27081(30) | 19 | - | 56 | - | 5 |
| 446 | 12011 | 01101 | 4041.393297(50) | 50 | 58 | - | 17 | 0.7 |
| 446 | 13312 | 02202 | 4044.72530(41) | 27 | 61 | - | - | 6 |
| 446 | 13311 | 02201 | 4044.72566(30) | 25 | 60 | - | - | 6 |
| 446 | 12211 | 01101 | 4053.694338(28) | 62 | 26 | - | 59 | 0.7 |
| 446 | 12212 | 01102 | 4053.694361(20) | 64 | 27 | - | 63 | 0.6 |
| 446 | 11111 | 00001 | 4061.9795930(58) | 111 | 66 | - | 45 | 0.3 |
| 456 | 12011 | 01102 | 3981.33317(18) | 39 | - | 46 | - | 6 |
| 456 | 12011 | 01101 | 3981.33378(16) | 49 | 63 | - | 34 | 6 |
| 456 | 11112 | 00001 | 3998.571578(94) | 47 | - | 64 | - | 6 |
| 456 | 11111 | 00001 | 3998.57184(23) | 57 | 47 | - | 34 | 8 |
| 546 | 12011 | 01101 | 4000.35765(22) | 41 | 43 | - | 29 | 6 |
| 546 | 12011 | 01102 | 4000.35946(50) | 16 | - | 33 | - | 4 |
| 546 | 11112 | 00001 | 4022.28933(29) | 38 | - | 66 | - | 7 |
| 546 | 11111 | 00001 | 4022.29036(16) | 53 | 63 | - | 14 | 7 |
| 448 | 11111 | 00001 | 4014.33116(20) | 54 | 62 | - | 10 | 7 |
| 448 | 11112 | 00001 | 4014.33169(29) | 30 | - | 57 | - | 6 |

[a] Isotopologues: 446: $^{14}N_2^{16}O$, 456: $^{14}N^{15}N^{16}O$, 546: $^{15}N^{14}N^{16}O$, 448: $^{14}N_2^{18}O$

Table 1: N$_2$O bands observed in the present TRFT-ICLAS spectra between 3901 and 4050 cm$^{-1}$. Energy level notation is $v_1\ v_2\ l\ v_3\ x$. The $v$'s and $l$ are the usual vibrational quantum numbers. $x = 1$ or $2$ stand for the usual $e$ and $f$ notations. Numbers in parentheses after the band center values give one standard deviation in units of the least significant digits. N is the total number of observed lines. RMS : Root Mean Square of the polynomial fit.

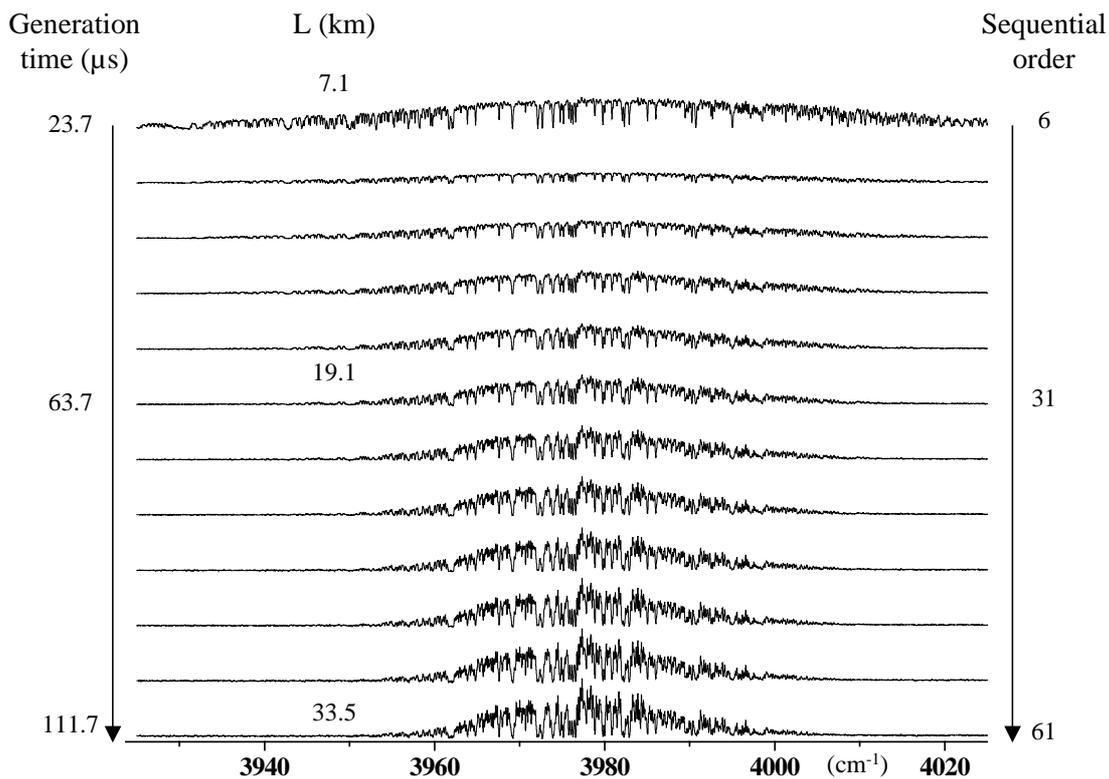

Figure 1: $N_2O$ time-resolved spectrum made of 64 time-components. Only one time-component out of five is plotted, starting from time-component n° 6. On the plot, two consecutive components, at an intermediate 0.071 cm$^{-1}$ apodized resolution, are 8 μs from each other. This corresponds to a 2.4-kilometer increase of the equivalent absorbing path L. $N_2O$ pressure is 70.58 hPa

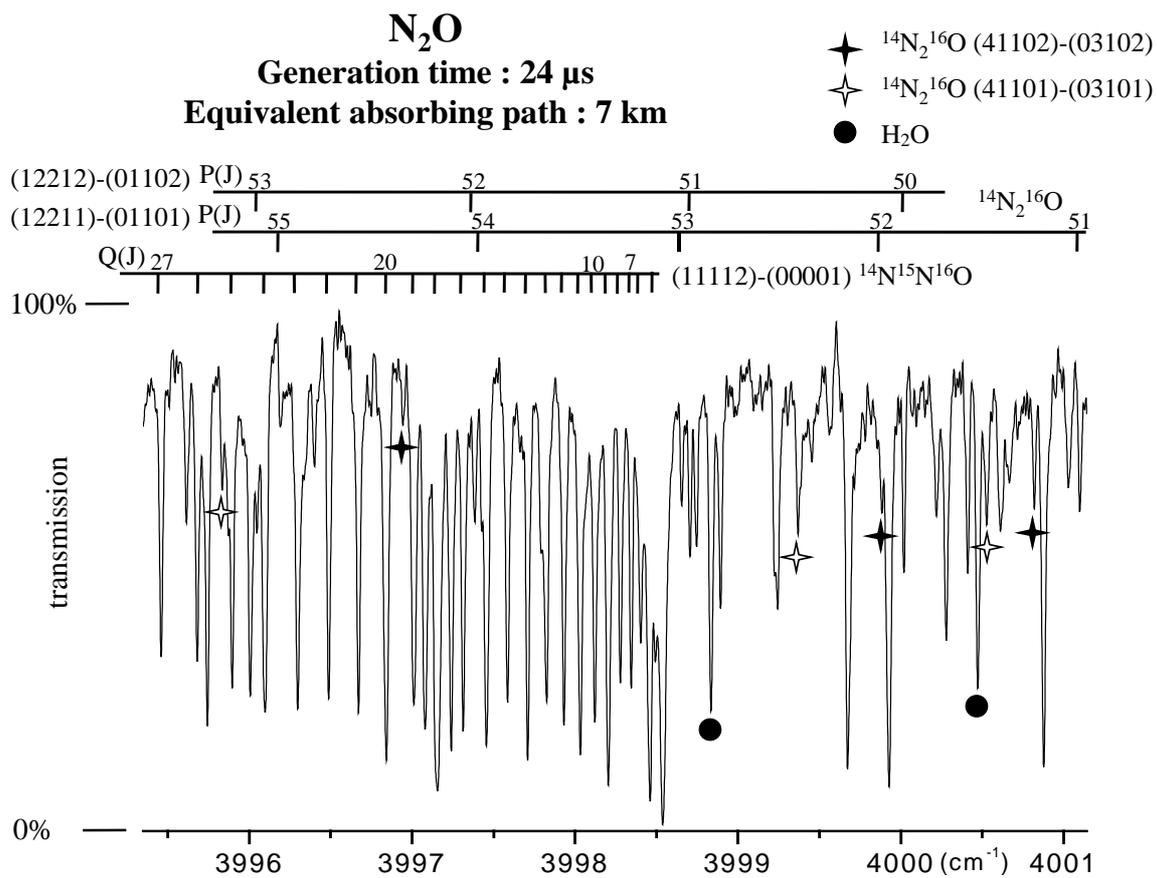

Figure 2: Limited portion of the time-component n° 6 shown on Figure 1. Spectral resolution is 7.1 $10^{-3}$ cm$^{-1}$. For the sake of clarity all assignments are not given.